\documentclass[twocolumn,letterpaper]{IEEEAerospaceCLS}  

\usepackage{graphicx}    
\usepackage{amsmath}
\usepackage{amsfonts}
\usepackage{bm}
\usepackage{hyperref}

\newcommand{\ignore}[1]{}

\begin{document}
\title{Generative Modeling of Microweather Wind Velocities for Urban Air Mobility }

\author{%
Tristan A. Shah\\ 
NASA Langley Research Center\\
Hampton, VA 23681 \\
tristan.a.shah@nasa.gov
\and 
Michael C. Stanley\\
Analytical Mechanics Associates \\
Hampton, VA 23666 \\
michael.c.stanley@nasa.gov
\and 
James E. Warner\\
NASA Langley Research Center \\
Hampton, VA 23681 \\
james.e.warner@nasa.gov
\thanks{\footnotesize 979-8-3503-5597-0/25/$\$31.00$ \copyright2025 IEEE}              
}

\maketitle

\thispagestyle{plain}
\pagestyle{plain}

\maketitle

\thispagestyle{plain}
\pagestyle{plain}

\begin{abstract}
Motivated by the pursuit of safe, reliable, and weather-tolerant urban air mobility (UAM) solutions, this work proposes a generative modeling approach for characterizing microweather wind velocities. 
Microweather, or the weather conditions in highly localized areas, is particularly complex in urban environments owing to the chaotic and turbulent nature of wind flows. 
Furthermore, traditional means of assessing local wind fields are not generally viable solutions for UAM applications: 1) field measurements that would rely on permanent wind profiling systems in operational air space are not practical, 2) physics-based models that simulate fluid dynamics at a sufficiently high resolution are not computationally tractable, and 3) data-driven modeling approaches that are largely deterministic ignore the inherent variability in turbulent flows that dictates UAM reliability. 
Thus, advancements in predictive capabilities are needed to help mitigate the unique operational safety risks that microweather winds pose for smaller, lighter weight UAM aircraft. 

This work aims to model microweather wind velocities in a manner that is computationally-efficient, captures random variability, and would only require a temporary, rather than permanent, field measurement campaign. 
Inspired by recent breakthroughs in conditional generative AI such as text-to-image generation, the proposed approach learns a probabilistic macro-to-microweather mapping between regional weather forecasts and measured local wind velocities using generative modeling. 
A simple proof of concept was implemented using a dataset comprised of local (micro) measurements from a Sonic Detection and Ranging (SoDAR) wind profiler along with (macro) forecast data from a nearby weather station over the same time period. 
Generative modeling was implemented using both state of the art deep generative models (DGMs), denoising diffusion probabilistic models and flow matching, as well as the well-established Gaussian mixture model (GMM) as a simpler baseline. 
Using current macroweather forecasted wind speed and direction as input, the results show that the proposed macro-to-microweather conditional generative models can produce statistically consistent wind velocity vs. altitude samples, capturing the random variability in the localized measurement region. 
While the simpler GMM performs well for unconditional wind velocity sample generation, the DGMs show superior performance for conditional sampling and provide a more capable foundation for scaling to larger scale measurement campaigns with denser spatial/temporal sensor readings. 

\end{abstract}

\tableofcontents

\section{Introduction} \label{sec:intro}

\begin{figure*}
    \centering
    \includegraphics[width=\linewidth]{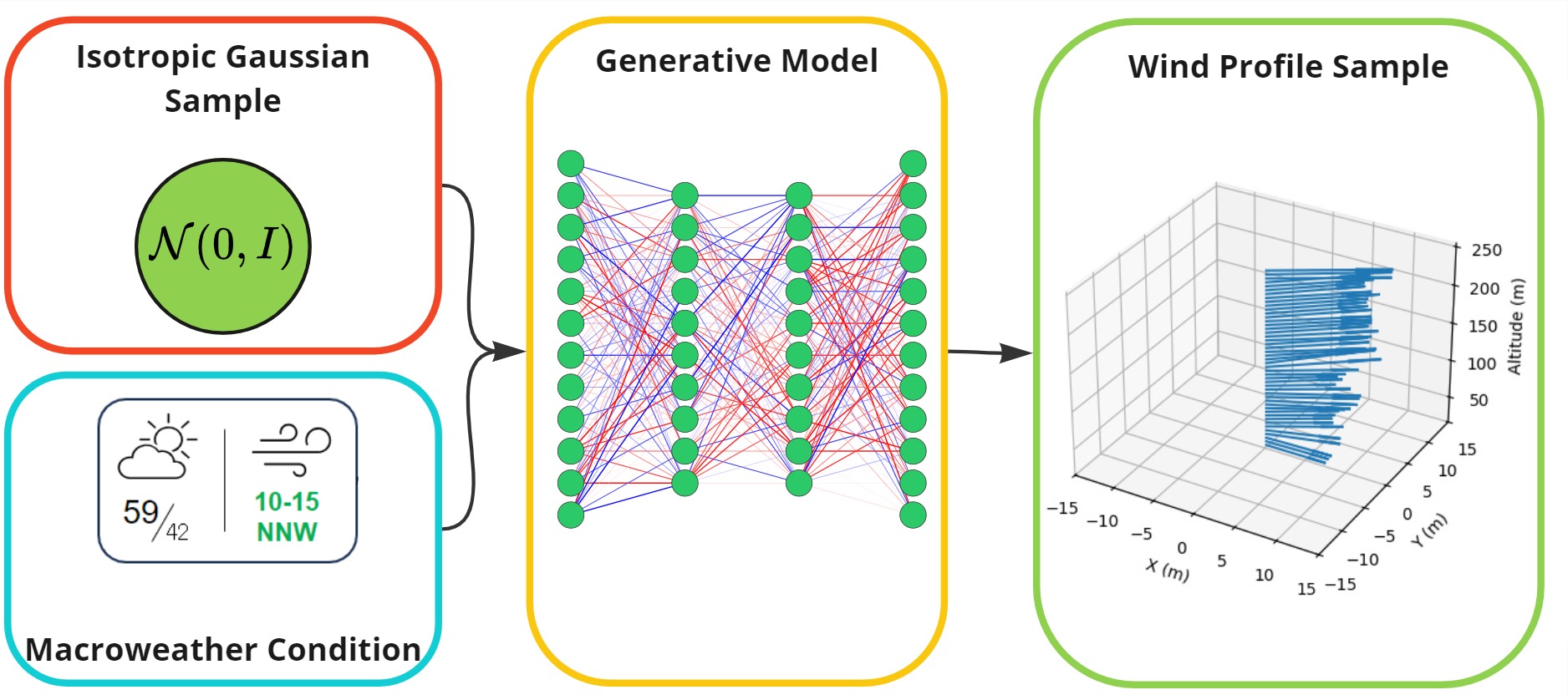}
    \caption{\textbf{Our method for utilizing conditional deep generative models to produce distributions over wind profiles based on known macroweather conditions.}}
    \label{fig:main_scheme}
\end{figure*}

Urban air mobility (UAM) has the potential to revolutionize the transportation of people and cargo in small and often autonomous aircraft that can circumvent increasingly congested ground infrastructure \cite{faa_uam}.
While recent advances in electrification, automation, and vertical take-off and landing (VTOL) capabilities have accelerated progress towards commercial viability, several challenges related to noise, air traffic management, infrastructure, and weather still prevent the mainstream adoption of UAM solutions \cite{cohen_2021}.
Of these challenges, the ability to assess weather conditions in localized urban environments (\textit{microweather}) on short time horizons (\textit{nowcasting}) is critical because of its technical complexity and its significant impact on the safety, reliability, and weather tolerance of UAM operations.

This work considers the microweather nowcasting problem in the context of assessing local wind velocity fields. 
Wind estimation is particularly complex owing to the chaotic and turbulent nature of flow fields through urban terrains. 
Microweather winds also pose many unique operational safety risks such as trajectory deviation, rapid deterioration of battery charge, and decreased passenger comfort for UAM \cite{drones8040147} since the sensitivity of aircraft and passengers increase with decreasing aircraft size \cite{cohen_2021}. 
These risks are especially prevalent during takeoff, landing, and transition phases of flight due to the close proximity with buildings. Despite the importance of microweather nowcasting for winds, this capability largely lags behind meso/macro weather forecasting where the resolution is too coarse to have practical use for UAM.

Attempts to assess local wind velocity fields have traditionally been performed using either experimental measurements or physics-based modeling. 
The experimental approach consists of either wind tunnel testing of downscaled domains \cite{UEHARA20001553} or full scale field measurements which can be performed using ground-based \cite{doi:10.2514/6.2023-2240,rs3091871} or airborne sensors \cite{zhengnong_2023,s20051341}. 
While field measurements are the most direct and reliable way to estimate microweather winds, the permanent installation of wind profiling systems is not practical in UAM areas of interest such as an operational takeoff/landing zone. 
Physics-based modeling approaches, most commonly using computational fluid dynamics (CFD), aim to numerically solve the Navier-Stokes equations governing fluid flow throughout the atmospheric boundary layer (ABL). 
Significant advancements in the ability to model complex and turbulent flow fields in the ABL have been made for UAM applications \cite{drones8040147,fluids7070246}, and have also been motivated by pedestrian comfort and ventilation \cite{en13112827,ANTONIOU2019133743} and renewable energy \cite{TOJASILVA201866} research. 
Still, it remains a challenge to accurately prescribe initial/boundary conditions needed for CFD models and the formidable computational expense of these approaches limit their applicability as a viable microweather nowcasting approach.

More recently, data-driven estimation of microweather wind velocities has become increasingly common as a computationally efficient alternative to physics-based modeling. 
Such approaches rely on large experimentally measured or simulated training datasets to construct machine learning or statistical models for local wind fields \cite{10.1063/5.0091980}. 
Earlier efforts focused on more traditional machine learning models like Gaussian process regression \cite{stock2018wind} while recent work is dominated by deep learning \cite{chrit_2024,s23073715,ZHANG2021116641,10592305}. 
A large majority of these studies posed wind estimation as a supervised learning task: make a \textit{deterministic} prediction of wind velocity (output) for given flow parameters (inputs). 
Ignoring the randomness in the estimation problem crucially prevents these approaches from capturing inherent variability in turbulent flows that strongly dictate UAM reliability. 
Studies using statistical methods like autoregressive and Markov models viewed winds probabilistically, but typically focus on modeling wind flows across larger meso/macro scales and/or in offshore or higher elevation areas \cite{hering2015markov,pinson2012adaptive,tagle2019non}.

The proposed work explores the use of generative modeling as a data-driven nowcasting technique that can capture random variability observed in wind velocity fields. 
Generative models learn to synthesize new samples from an unknown probability distribution using a sufficiently large training dataset of representative samples. 
Original attempts to do so relied on building explicit parametric approximations to the target probability distribution function. 
The most commonly used parametric approach is the Gaussian mixture model (GMM), which employs a weighted sum of Gaussian component densities \cite{bishopML} and uses the Expectation-Maximization (EM) \cite{em_orig} to fit its parameters (weights, mean vectors, covariance matrices). 
GMMs have been used successfully for decades on a range of applications, but can struggle with high-dimensional data and large training datasets.

In recent years, Deep Generative Models (DGMs) have surged in popularity for synthesizing complex and high dimensional signals including images, video, and audio.
DGMs are a family of neural network-based models that learn a mapping between a simple source distribution (e.g., isotropic Gaussian) and the target probability distribution. 
DGMs, in contrast to GMMs, make no assumption about the underlying structure of the distribution they learn and leverage the scalability of deep neural networks to accommodate potentially huge training datasets.

Early work on DGMs focused on directly mapping samples from the source to target distribution with a single neural network evaluation, enabling notably fast inference (i.e., sample generation) times. 
Variational Autoencoders (VAEs) \cite{Kingma2013AutoEncodingVB} and Generative Adversarial Networks (GANs) \cite{goodfellow2014} are approaches in this category, each with their own drawbacks. 
VAEs have been observed to generate samples with good coverage over the target distribution, however, individual samples appear to be empirically low quality and contain artifacts which allow them to be identified as ``artificial" \cite{xiao2021tackling}. 
In contrast, GANs have been shown to generate extremely high-quality samples such as images which are indistinguishable from real images \cite{brock2018large}. 
Unfortunately, GANs are difficult to train \cite{saxena2021generative} and often suffer from mode collapse, where only a small subset of the target distribution is learnt by the model.

Recent developments in DGMs have shown a convergence towards models that perform inference as a sequential process that results in improved sample quality. 
Denoising Diffusion Probabilistic Models (DDPMs) define a discrete time forward process by which any sample from a target distribution can be sequentially corrupted such that it becomes a member of the source distribution \cite{ho2020denoising}, \cite{nichol2021improved}. 
A reverse process is learned by the DDPM such that the model can iteratively move samples from the source to the target distribution. 
Similarly, Flow Matching (FM) defines a continuous time probability flow in which a time dependent velocity field carries samples from the source distribution to the target distribution \cite{lipman2022flow}. 
Due to their high sample quality \cite{rombach2022high}, coverage of the target distribution \cite{ho2020denoising}, and training stability \cite{ho2020denoising}, \cite{lipman2022flow}, sequential models such as DDPM and FM have been demonstrated as the state-of-the-art. 

There have been some recent attempts to leverage DGMs for modeling wind field variability. 
A large amount of this work has been motivated by renewable energy applications, where researchers have addressed the problem of probabilistic scenario generation for wind power using generative models \cite{DUMAS2022117871,dong2023short,ge2020modeling}. 
Conditional generative models have been used to predict wind flows using building geometries as inputs \cite{kastner2023gan,mokhtar2020conditional} and for super resolution methods that translate low-fidelity CFD solutions to high-fidelity predictions \cite{tran2020gans}. 
In these studies, however, the generative model acts mainly as a surrogate for expensive CFD models rather than focusing on modeling wind variability. 

A line of work that is more similar to the proposed approach here is flow reconstruction with generative models, where full velocity fields are probabilistically generated based on sparse measurements \cite{HU2024111120,10.1063/5.0172559,du2024confild}. 
In particular, one study \cite{10.1063/5.0172559} used diffusion models to reconstruct turbulent flows in ABLs with synthetic data from CFD. 
A primary challenge for flow reconstruction approaches is generating training data that authentically replicates real-world randomness via CFD simulation. 
Furthermore, implementing these approaches for UAM nowcasting would still require permanent measurement campaigns to provide the inputs for reconstructing flows in areas of interest.

This paper proposes an alternative approach for generative modeling of microweather wind velocities for UAM. 
Drawing inspiration from recent conditional generative AI breakthroughs such as text-to-image generation, a probabilistic mapping is learned from regional (macro) weather conditions to measured microweather wind conditions in an area of interest (e.g., a landing zone). 
Once this conditional generative model is trained, statistically realistic samples of local wind conditions can be generated on demand based on current weather forecasts to inform UAM operations. 
Such an approach would only require a temporary measurement campaign in an area of interest, circumventing the need for permanent sensor installations or the challenge of simulating realistic and probabilistic wind flow fields with CFD for training data.

A simple proof-of-concept of the approach was implemented using data from a recent NASA measurement campaign that collected wind velocity versus altitude data via a Sonic Detection and Ranging (SoDAR) wind profiler. 
Macroweather forecast data from a nearby weather station was obtained as conditioning information during the same time frame. 
DDPMs and flow matching, representing the current state of the art in generative AI, were compared to GMMs as a well-established and simpler baseline generative modeling approach. 
All three methods were shown to generate statistically consistent wind velocity samples based on the SoDAR measurement data, with the DGMs demonstrating superior performance in the conditional modeling setting.

To the best of the authors' knowledge, the proposed approach represents the first attempt to build such a probabilistic macro-to-microweather mapping with generative modeling and the first application of flow matching for modeling wind flow variability. 
While the concept is demonstrated in a simplified setting with relatively sparse and low-dimensional measurements, the DGM algorithms tested are capable of scaling to much larger datasets with a higher density of sensor measurements and larger numbers of macroweather conditions. 
Furthermore, this work identifies avenues for future algorithmic improvements to interpolate/extrapolate beyond sparse sensors to realize a framework more practical for UAM.

The remainder of the paper is organized as follows: first, \autoref{sec:background} provides a background on generative modeling, introducing the GMM, DDPM, and flow matching techniques that are adopted in this work. 
Then, the specifics of the macro-to-microweather generative modeling approach are given in \autoref{sec:approach}, including a description of the datasets used to implement a proof of concept. 
\autoref{sec:results} then compares the generated wind velocities using the proposed DGM approach with a baseline modeling technique and withheld test data. 
Finally, conclusions and potential avenues for future work are discussed in \autoref{sec:conclusions}.

\section{Background} \label{sec:background}
In this section technical details on the explored generative algorithms (GMM, DDPM, and FM) are provided.
Generally, the problem generative models solve is to learn the underlying probability density function $q(\bm{x})$ of random vector, $\bm{x} \in {\mathbb{R}}^d$, from which a finite number of data samples that have been observed $\bm{x}^{(1)}, \bm{x}^{(2)}, \cdots, \bm{x}^{(n)}$.
Many generative models take the approach of approximating $q(\bm{x})$ with a simpler, parameterized density $p(\bm{x})$.
A common approach to represent a parameterized density is through a latent variable formulation $p(\bm{x})=\int_z p(\bm{x}|\bm{z})p(\bm{z})d\bm{z}$.
In this formulation, simple conditional distributions $p(\bm{x}|\bm{z})$ are marginalized over a latent variable $\bm{z}$ to produce a more complex distribution capable of representing $q(\bm{x})$. 
Different choices in the implementation of latent variable models lead to a broad family of generative models.

\subsection{Gaussian Mixture Models} \label{sec:gmm_background}
The Gaussian Mixture Model (GMM) is a classical latent variable model for density estimation, and thus provides an effective baseline comparison for the DGMs.
The GMM posits that the data distribution can be approximated by $p(\bm{x})$, a convex combination of Gaussian distributions,
\begin{equation} \label{eq:gmm}
	p(\bm{x}) = \sum_{k = 1}^K \pi_k {\mathcal{N}}\left( \bm{x} \mid \bm{\mu}_k, \bm{\Sigma}_k \right),
\end{equation}
where $\pi_k \in [0, 1]$ is the weight of the $k$th component, $\bm{\mu}_k \in {\mathbb{R}}^d$ is the expectation of the $k$th component and $\bm{\Sigma}_k \in {\mathbb{R}}^{d \times d}$ is the covariance of the $k$th component.
Additionally, it must hold that $\sum_k \pi_k = 1$.
As described in \cite{bishopML}, GMM's can be viewed as a description of the data generating process involving a latent variable which first randomly chooses the component of the GMM and then randomly draws from that component's Gaussian distribution.
Although this perspective does not have an obvious motivation in the application considered herein, it is useful when thinking about how to draw samples from a GMM and underscores the ease of generating data from a GMM.
We further emphasize that although GMM's are the convex combination of Gaussian distributions, they are able to express highly non-Gaussian distributions, adding to their suitability as a baseline model comparison.

Given data $\bm{x}^{(1)}, \bm{x}^{(2)}, \dots, \bm{x}^{(n)} \sim q(\bm{x})$, where $\bm{x}^{(i)} \in {\mathbb{R}}^d$ for all $i$, the task of fitting the GMM as defined by Equation \eqref{eq:gmm} is equivalent to choosing the number of components, $K$, and estimating the parameters $(\pi_k, \bm{\mu}_k, \bm{\Sigma}_k)$ for $k = 1, \dots, K$.
Since for each component, a full covariance matrix adds $\frac{1}{2}(d^2 + d)$ parameters to estimate, one often opts to make structural assumptions about the covariance matrix (e.g., diagonal covariances or one covariance matrix for all components).
We retain the full covariance structure to ensure that each model component is capturing covariance across wind velocities at different altitudes. 

For a fixed $K$, the EM algorithm is typically used to estimate the parameters by iterating between setting component weights and updating the maximum likelihood estimates of the Gaussian parameters given the component weights \cite{em_orig}, \cite{bishopML}. 
To select the number of components, $K$, one typically evaluates the Bayesian Information Criterion (BIC) \cite{bic_schwarz}, \cite{esl} across different choices of $K$.
The BIC takes the form,
\begin{equation}
	\text{BIC}(k) = -2 \log L(k) + \phi \log n,
\end{equation}
where $L(k)$ is a shorthand for the likelihood of the data under a GMM with $k$ components and $\phi$ is the number of parameters to estimate for a $k$-component GMM.
More specifically, there are $1 + d + \frac{1}{2}(d^2 + d)$ parameters to estimate per component and hence, $\phi = k(1 + d + \frac{1}{2} (d^2 + d))$.
Thus, the BIC provides a quantitative measure of model fit as measured by the log-likelihood versus the model complexity as measured by the number of parameters and number of data points.
The choice of $K$ can be made by plotting the BIC as a function over a grid of $k$ and choosing the point where the log-likelihood improvement no longer eclipses the increasing model complexity (as measured by $\phi \log n$).

Once a satisfactory GMM fit has been made to the data, the GMM can be used to sample realizations from the full joint distribution.
This capability further allows easy sampling from some conditional distributions via rejection sampling.
Let $GMM(K)$ denote the fitted model with $K$ components and suppose we draw the following samples from the full joint distribution, $\bm{z}^{(1)}, \bm{z}^{(2)}, \dots, \bm{z}^{(m)} \sim GMM(K)$, where $\bm{z}^{(i)} \in {\mathbb{R}}^d$ for all $i$.
Suppose we are interested in the conditional distribution of  the $2:d$ components of $\bm{z}^{(i)}$ given that the first element is contained in a set, $\mathcal{C} \subset \mathbb{R}$.
Samples from this conditional distribution can be obtained by simply ignoring those samples $\bm{z}^{(i)}$ such that $\bm{z}^{(i1)} \notin \mathcal{C}$.
The remaining samples are drawn from the correct conditional distribution.
We use this capability in \autoref{sec:results} to sample from the distributions of wind speed conditional on macroweather conditions.
See \autoref{sec:gmm_implement} for details on our GMM implementation.

\subsection{Denoising Diffusion Probabilistic Models}

DDPMs \cite{ho2020denoising} are a type of generative model that breaks sample generation into a sequential process.
In the DDPM framework, a discrete time process $\bm{x}_0, \cdots, \bm{x}_T$ with $T$ timesteps is defined where data at the final timestep $\bm{x}_T \sim {\mathcal N}(0, I)$ is a tractable distribution such as an isotropic Gaussian.
Reversing the process from $t=T$ to $t=0$ generates the target distribution $\bm{x}_{0} \sim q(\bm{x}_{0})$ from which samples are available. 
Interestingly, DDPMs can be considered to be latent variable models \cite{ho2020denoising} where the marginalization is performed over trajectories $p(\bm{x}_0) := \int p(\bm{x}_{0:T}) d\bm{x}_{1:T}$.
A distinguishing feature of DDPMs is that their forward process is explicitly defined by gradually corrupting a sample from the data distribution with Gaussian noise until it reaches the final timestep. 
Fortunately, the forward process can be written in closed form in Equation \eqref{eq:corrupt} which provides convenient access to a corrupted sample at a particular timestep. 
In Equation \eqref{eq:corrupt} $\bm{\epsilon} \sim {\mathcal N}(0, I)$ is Gaussian noise which corrupts the data $\bm{x}_{0}$ and $\bar{\alpha}_t$ is a time dependent parameter which controls the rate of data corruption. 
The rate at which $\bar{\alpha}_t$ changes is determined by a noise scheduler function, which is a hyperparameter of the DDPM algorithm. 
Many schedulers have been proposed in order to improve DDPM's performance \cite{nichol2021improved}, however in this work we employ the simplest (linear) scheduler.

\begin{equation}\label{eq:corrupt}
    \bm{x}_t := \sqrt{\bar{\alpha}_t} \bm{x}_0 + \sqrt{1 - \bar{\alpha}_t}\bm{\epsilon}
\end{equation}
Training the DDPM algorithm is relatively straightforward and can be accomplished by minimizing Equation \eqref{eq:diffusion_loss}.
In order to minimize this expectation, the neural network $\epsilon_\theta$ must learn to identify the noise which is present in a corrupted data sample.
Specifically, the expectation is minimized over data sampled from the training dataset, $p(\bm{x}_0)$, and timesteps sampled uniformly. 

\begin{equation}\label{eq:diffusion_loss}
    L(\theta) = {\mathbb E}_{\bm{x}_0 \sim p(\bm{x}_0), t \sim U(0, T)} [||\bm{\epsilon} - \epsilon_{\theta}(\bm{x}_t, t)||^2]
\end{equation}

After $\epsilon_\theta$ has been trained it can be utilized to generate novel samples from $p(\bm{x}_0)$ by reversing the forward diffusion process through an iterative sampling procedure.
Starting with a random sample from the tractable source distribution: $x_T$, noise is removed by applying Equation \eqref{eq:diffusion_sample} until $\bm{x}_0$ is reached \cite{ho2020denoising}.
In this equation, $\bm{z}$ is gaussian noise weighted by $\sigma_t$ which decays to $0$ as $t \rightarrow 0$, allowing for stochastic denoising trajectories.

\begin{equation}\label{eq:diffusion_sample}
    \bm{x}_{t-1}=\frac{1}{\sqrt{\alpha_t}}\big(\bm{x}_t-\frac{1-\alpha_t}{\sqrt{1-\bar{\alpha}_t}}\epsilon_\theta(\bm{x}_t, t)\big) + \sigma_t\bm{z}
\end{equation}

\subsection{Flow Matching}

The Flow Matching (FM) algorithm shares similarities with DDPM in both motivation and implementation \cite{lipman2022flow}.
Similar to DDPM, FM is a latent varaible model which marginalizes over trajectories, however, it operates in continuous $t\in[0, 1]$ instead of discrete time.
The main idea behind FM is to learn a time dependent probability flow Equation \eqref{eq:continuity} which carries probability mass from a tractable (e.g., Gaussian) source distribution\footnote{In FM literature the notation for source and target distribution differ from DDPM literature which uses $t=0$ for the target and $t = T$ for the source.}, $\bm{x}_0\sim p_0(\bm{x})$, to a target distribution, $\bm{x}_1\sim p_1(\bm{x})$.
Previous works considered learning the parameters of a flow using maximum likelihood estimation \cite{chen2018neural}, however, this is not a scalable objective to high dimensional problems.
FM introduced an efficient method by which a parameterized flow field $v_\theta(\bm{x}, t)$ can be learnt from samples of any source and target distribution \cite{lipman2022flow} \cite{tong2023improving}.

\begin{equation}\label{eq:continuity}
    \frac{\partial p(\bm{x})}{\partial t} = -\nabla \cdot (p_t(\bm{x}) v_\theta(\bm{x}, t))
\end{equation}

The flow field can be learned by first constructing conditional probability paths between pairs of samples from the source and target distributions.
In our work we employ the choices for mean and variances of the paths introduced by \cite{tong2023improving}. 
The mean of these paths is chosen as a linear interpolation between a source and target sample $\bm{\mu}_t := t\bm{x}_1 + (1 - t)\bm{x}_0$ while the standard deviation $\sigma$ is set to a small positive constant. 
Now, $v_\theta$ can be learned by minimizing Equation \eqref{eq:fm_loss}
\begin{equation}\label{eq:fm_loss}
    L(\theta) = {\mathbb E}_{\bm{x}_0\sim p_0, \bm{x}_1\sim p_1,t\sim U(0, 1)}[||v_\theta(\bm{x}_t, t) - (\bm{x}_1 - \bm{x}_0)||^2]
\end{equation}
where $\bm{x}_t \sim {\mathcal N}(\bm{\mu}_t, \sigma)$.

It has been shown \cite{lipman2022flow,tong2023improving} that minimizing Equation \eqref{eq:fm_loss} results in a $v_\theta$ which produces optimal transport paths between $p_0$ and $p_1$. 
These paths are straighter \cite{tong2023improving} than those produced by optimizing Equation \eqref{eq:diffusion_loss} and therefore require fewer function evaluations to generate samples.
Generating samples from $p_1$ can be achieved through numerical integration of Equation \eqref{eq:fm_sample}
\begin{equation}\label{eq:fm_sample}
    \bm{x}_1 = \bm{x}_0 + \int_{0}^{1}v_\theta(\bm{x}_t, t)dt
\end{equation}
where a coarser time discretization can be used to trade off sample quality for computation speed.

\section{Approach} \label{sec:approach}
In this work our goal is to leverage recent advances in generative modeling to learn a probabilistic map between macroweather and microweather as illustrated in \autoref{fig:main_scheme}. 
To accomplish this goal we train generative models on a dataset of macro-micro weather pairs.
In this section we discuss details regarding the dataset as well as the implementation details of our models.

\subsection{Dataset}

\begin{figure}
    \centering
    \includegraphics[width=\linewidth]{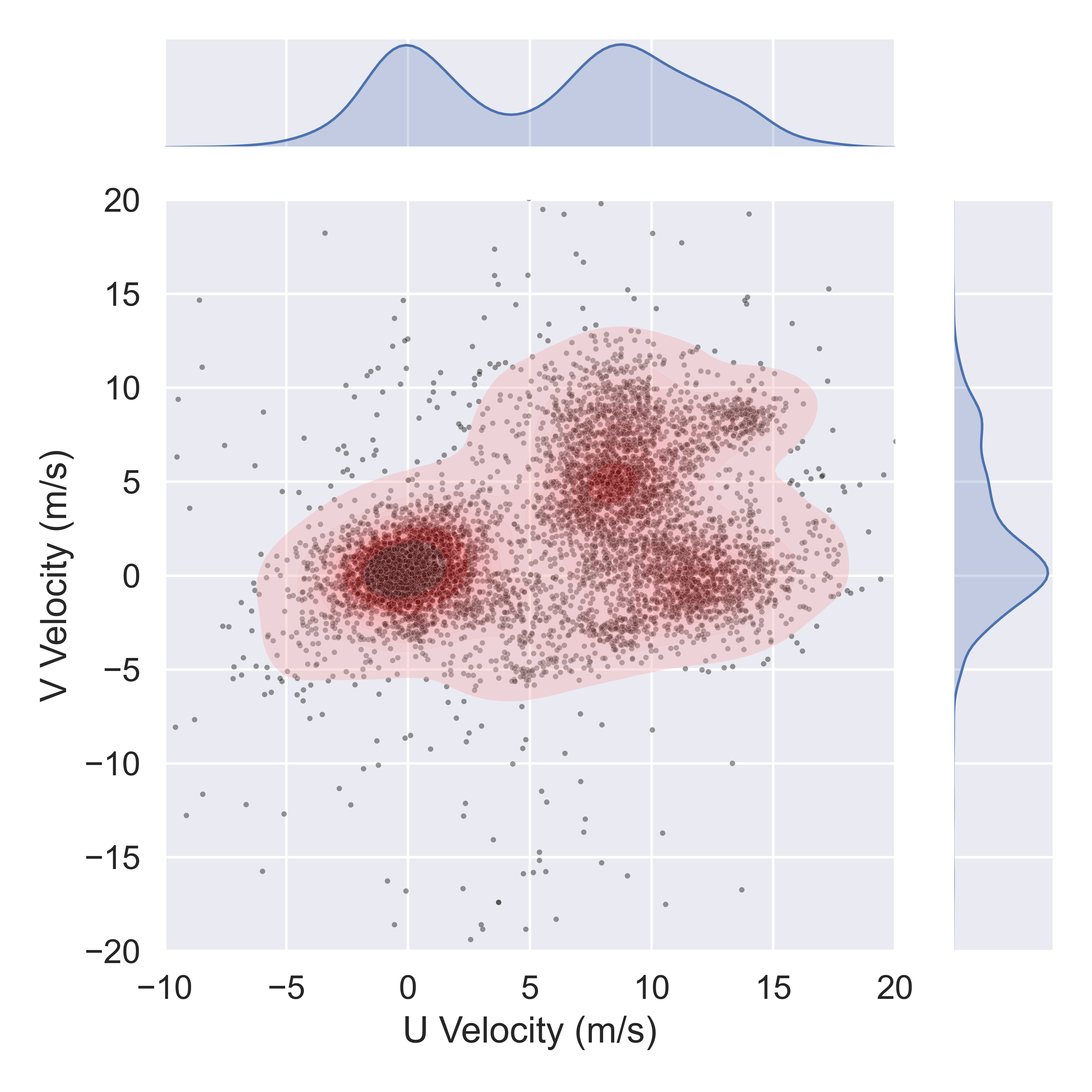}
    \caption{\textbf{Empirical distribution of SoDAR-measured velocity components averaged over altitudes. This plot represents all 6542 observations in the dataset.}}
    \label{fig:real_bivariate}
\end{figure}

Wind profiles provide useful information about how wind velocity varies as a function of space at a particular time. 
In this work, we consider wind profiles collected through SoDAR sensors \cite{coulter2020sonic}, which estimate wind velocity components by measuring back-scattering of an acoustic pulse through the atmosphere.
A measurement campaign at NASA Armstrong Flight Research Center (AFRC) collected SoDAR measurements in 2 minute intervals over the course of 10 days (4/21/22 - 4/30/22) at 47 evenly-spaced altitudes from 20m to 250m, yielding 6,542 observations.
The corresponding macroweather forecast for the region was collected from the nearby weather station on Edwards Air Force Base. 

\begin{figure*}
    \centering
    \includegraphics[width=\textwidth]{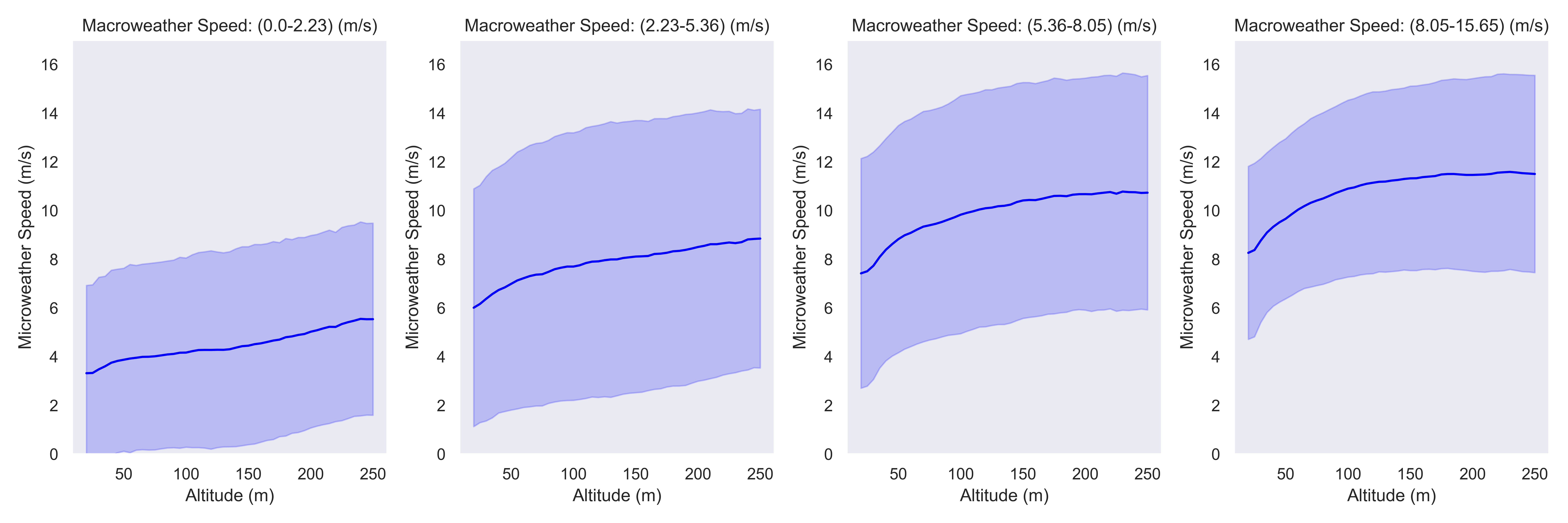}
    \caption{\textbf{Microweather wind speed altitude profiles under different macroweather wind speed conditions where the mean wind speed over altitude is plotted in dark blue with plus and minus one standard deviation shown by the shaded blue regions. Each macroweather wind speed range contains an approximately equal number of samples. High macroweather wind speeds are associated with higher microweather wind speeds.}}
    \label{fig:macro_micro_speed}
\end{figure*}

Our dataset is constructed as a set of tuples:
\begin{equation*}
    {\mathcal{D}} = \left\{\bm{x}^{(i)}, \bm{c}^{(i)} \right\}_{i = 1}^N,
\end{equation*}
where $\bm{x}^{(i)} \in {\mathbb{R}}^d$ is a vector of altitude velocity components and $\bm{c}^{(i)}$ is a vector of variables describing macroweather conditions. 
Namely, each of the 47 altitudes has a $u \in \mathbb{R}$ and $v \in \mathbb{R}$  component and thus, $d = 47 \cdot 2 = 94$. The conditioning variable $\bm{c}^{(i)}$ can be encoded as either a vector of categorical or numerical elements. 
In this work we consider macroweather conditions of forecasted wind direction and speed. 
If using categorical elements, direction is one of 16 compass directions and wind speed is split into four bins as shown in the subplot titles in \autoref{fig:macro_micro_speed}. 
If using numerical elements, direction and wind speed can be transformed into $u$ and $v$ velocity components.

In this study we select $\bm{c}$ to be made up of two pieces of macroweather information: direction and speed.
In principle, however, any data source that is correlated with the microweather wind velocities and can be conveniently accessed into the future can act as effective conditioning variables for the proposed approach. 
For example, $\bm{c}$ could represent wind measurements from a simple instrument like an anemometer at a peripheral location, provided it is close enough to provide useful signal for a conditional generative model.
Due to the limited size of our dataset, temporal dependence in the data is removed in this work, viewing each sample of wind velocity in the training dataset as an independent measurement.
We defer to future work to explore larger datasets which contain enough measurements to generate wind fields both throughout space and over time.

The distribution of altitude-averaged microweather velocity components can be visualized in \autoref{fig:real_bivariate}. 
In this figure it is clear that the wind profiles form a multi-modal distribution with the largest mode centered around zero velocity. 
Higher wind velocities tend to cluster into the other two modes of the distribution. 
This figure demonstrates complexity in the altitude-averaged microweather distribution where no macroweather conditions are considered. 
A connection between macroweather speed and microweather wind profiles can be visualized in \autoref{fig:macro_micro_speed}. 
This figure demonstrates microweather distributions $p({\mathbf x}|{\mathbf c})$ conditioned on categorical macroweather speed ranges. 
As the macroweather speed range increases the microweather wind profile distribution also increases. 
Interestingly, for lower macroweather speeds the microweather profile appears approximately linear whereas the higher macroweather speeds show a characteristic logarithmic pattern. 
In the following \autoref{sec:results} we demonstrate each models' ability to capture these complexities in both the unconditional and conditional cases.

\subsection{GMM implementation} \label{sec:gmm_implement}
As mentioned in \autoref{sec:gmm_background}, the number of GMM parameters to estimate depends primarily on the chosen number of components, $K$ and the covariance matrix structure of each component.
Given the structure of the data as described above, we use a full covariance matrix to help each component capture the covariance between different altitude profiles.
With this choice, the number of parameters quickly expands beyond the number of observations as more GMM components are included.
For the GMM portion of the modeling, we concatenate the micro and macro conditions, yielding 96 total wind velocity elements per observation (47 $u$ micro elements, 47 $v$ micro elements, one $u$ macro element and one $v$ macro element).
As such, since each component adds 4,753 parameters to estimate, we propose a prepossessing data reduction step using principal component analysis (PCA) since the distribution is already unidentifiable at two components.

\begin{figure}
    \centering
    \includegraphics[width=0.75\linewidth]{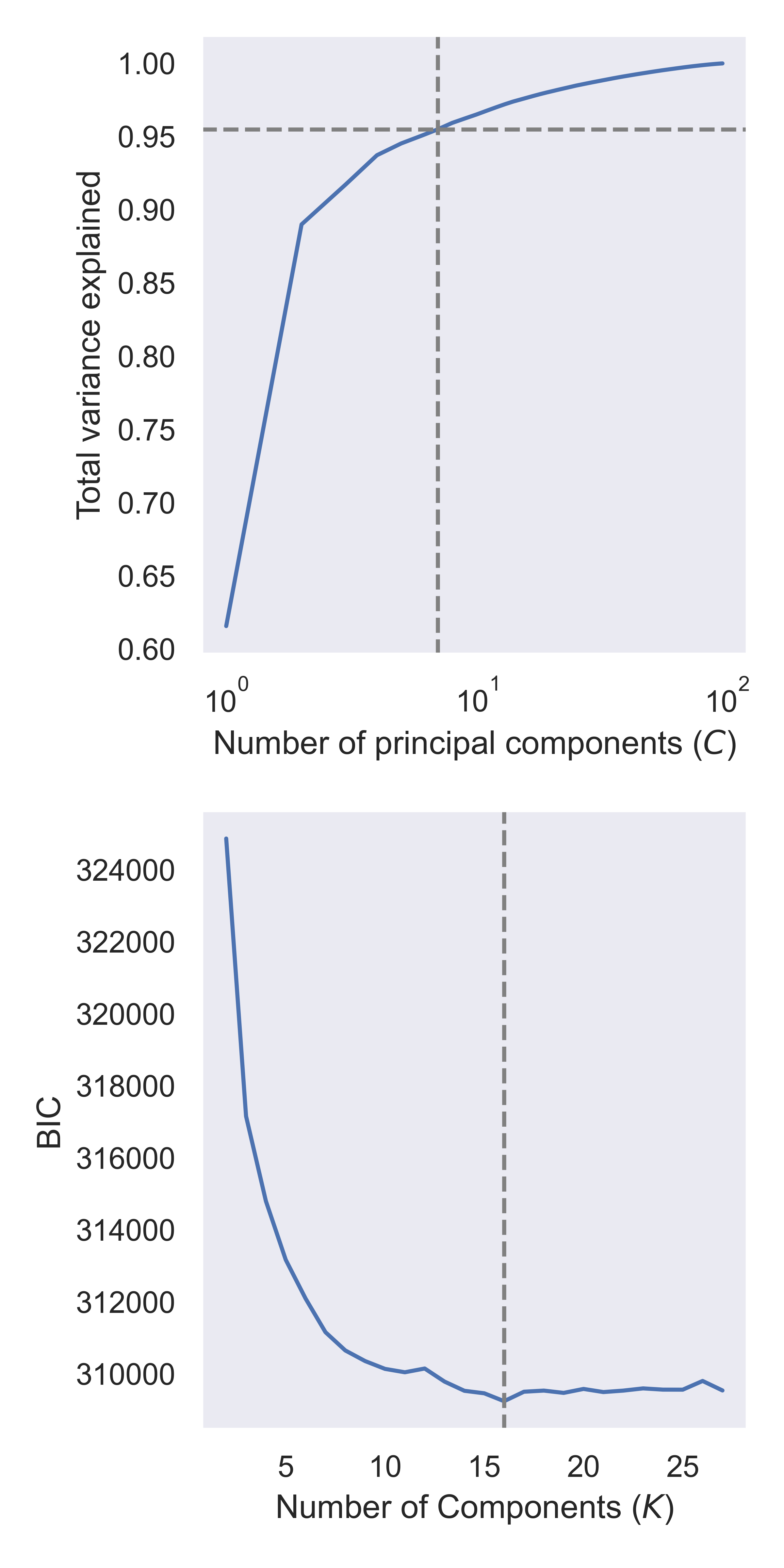}
    \caption{\textbf{Total variance explained for the PCA dimension reduction and BIC evaluation across the number of GMM components. Seven principal components explains $\approx 96.4\%$ of the variance while still achieving realistic results (see \autoref{sec:results}. $21$ GMM components is roughly the lowest model complexity achieving a low BIC.}}
    \label{fig:gmm_diagnostics}
\end{figure}

Let ${\mathbf{X}} \in {\mathbb{R}}^{N \times d'}$ denote the centered design matrix of the $N$ SoDAR observations where $d' = 94 + 2$.
To obtain a low-dimensional projection of these data, we first compute the eigendecomposition, ${\mathbf{X}}^T {\mathbf{X}} = {\mathbf{U}} {\mathbf{\Sigma}} {\mathbf{U}}^T$, where the columns of ${\mathbf{U}}$ are the eigenvectors of ${\mathbf{X}}^T {\mathbf{X}}$.
Let ${\mathbf{U}}_C \in {\mathbb{R}}^{C \times d'}$ denote the matrix composed of the first $C$ eigenvectors, as ordered by their eigenvalues in ${\mathbf{\Sigma}}$.
We project our data to the $C$-dimensional subspace defined by the eigenvectors via the linear transformation, ${\mathbf{Y}} = {\mathbf{X}} {\mathbf{U}}_C$, where each datum is now of the form $\bm{y}^{(i)} \in {\mathbb{R}}^C$.
A GMM is then fit to the $N$ $C$-dimensional observations where the BIC is used to choose the number of components.
To sample data from the original space, we first obtain a sample $\bm{z} \in {\mathbb{R}^C}$ from our fitted GMM and then map it back to the original space by the affine transformation, $\bm{x} = {\mathbf{U}}_C^T \bm{z} + \bar{\bm{x}}$, where $\bm{x} \in {\mathbb{R}}^{d'}$ is a vector of the column means from the original design matrix, ${\mathbf{X}}$.

To choose $C$, we considered the cumulative explained variance to select a reasonable cutoff.
With $7$ eigenvectors, we explain $\approx 96.4 \%$ of the variance and further found this choice to be a good balance between a low-dimensional computationally feasible representation and high-fidelity results.
Both explained variance as a function of principal components and the BIC as a function of the GMM components are shown in \autoref{fig:gmm_diagnostics}.
Twenty one GMM components were chosen since this is roughly the lowest complexity model obtaining minimum BIC results.

\subsection{Neural Network Implementation}
In our work, both the noise estimator $\epsilon_\theta(x_t, {\mathbf c}, t)$ for DDPM and the flow field $v_\theta(x_t, {\mathbf c}, t)$ for FM are realized by a U-net neural network architecture \cite{ronneberger2015u}.
The macroweather condition ${\mathbf c}$ is provided to the networks in addition to time $t$ (time in the sequential generation process).
The loss functions in Equation \eqref{eq:diffusion_loss} and Equation \eqref{eq:fm_loss} can be easily updated to include the additional ${\mathbf c}$ term. 
Additionally, when sampling using Equation \eqref{eq:diffusion_sample} and Equation \eqref{eq:fm_sample} ${\mathbf c}$ can also be provided.
Both ${\mathbf c}$ and $t$ were injected into the U-net at every layer as described in previous work \cite{rombach2022high} to provide a strong signal to the network.
This architecture was originally developed for medical image segmentation but has proven to be successful in generative modeling and is used as the neural network backbone in many state of the art generative systems \cite{ho2020denoising,rombach2022high}.
The U-net architecture is symmetric, in that its input and output tensors are the same dimensions. 
Internally, the U-net takes in input images and performs multiple downsampling operations followed by an equal number of upsampling operations that return the data to it's original shape.
A key feature of U-net are its residual connections \cite{he2016deep}, which are used across every stage of down and upsampling in order to improve training stability.
We adapt the original 2D U-net architecture to our wind profile data by changing the 2D convolutional layers to 1D convolutions.
This simple modification allows our networks to exploit the spatial relationship between velocities at nearby altitudes and to consider $u$ and $v$ velocity components as separate channels.

The code used to train the generative models and produce results for this work has been made publicly available on GitHub\footnote{See https://github.com/nasa/wind-generative-modeling.}
The GMM implementation which was used in this work is from the scikit-learn package in Python. 
Our 1D U-net and the DDPM and FM algorithms were implemented in Python using PyTorch and are included in the code base.
In the next section we demonstrate the results of applying these models to the task of generative modeling of microweather wind profiles.

\section{Results} \label{sec:results}

In this section we evaluate the ability of the three generative models intoduced in section \ref{sec:background} to model the microweather data distribution at different levels of complexity.
First, we show their ability to match the data distribution in the unconditional case without considering macroweather information. 
We show unconditional sample generation in an altitude-averaged case and then as a function of altitude.
Next, we demonstrate the quality of distribution produced by each model when conditioned on macroweather speed and direction. 
Finally, the capability of each model to generate samples on a previously unseen combination of macroweather conditions is investigated.

Comparisons of the models are evaluated quantitatively in terms of empirical symmetrized Kullback-Leibler (KL) divergence \cite{wang_kld,wang_kld_2}, as well as qualitatively by the appearance of their generated samples. Here, we emphasize the probabilistic nature of our proposed approach, where the output of the models considered is a probability distribution (or, in particular,  samples thereof) rather than a deterministic prediction of any single wind measurement. Thus, the KL divergence, a statistical measure of the difference between two probability distributions, is selected as our quantitative metric rather than the error between pairwise wind samples.

\subsection{Unconditional Sampling}

\begin{figure*}
    \centering
    \includegraphics[width=\linewidth]{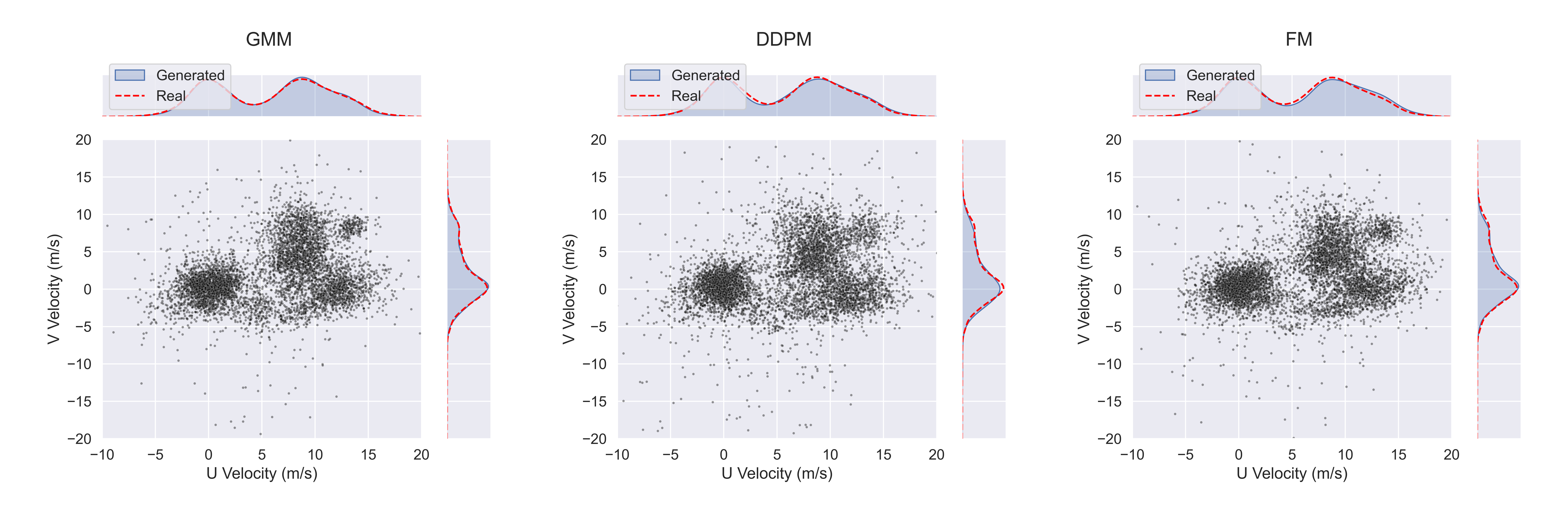}
    \caption{\textbf{Altitude-averaged: (left) Samples from the fitted GMM model along with kernel density estimations of the marginal $u$ and $v$ distributions. Both the joint and marginal distributions closely resemble those seen in \autoref{fig:real_bivariate}, providing strong evidence that the GMM is fitting the observed data distribution well. (middle) Samples from the fitted DDPM along with kernel density estimations of the marginal $u$ and $v$ distributions. This fit misses some of the visually identifiable modes as seen in \autoref{fig:real_bivariate}. (right) Samples from the fitted FM along with kernel density estimations of the marginal $u$ and $v$ distributions. Similar to those results in the DDPM bivariate plot, this fit misses some of the visually identifiable modes as seen in \autoref{fig:real_bivariate}}}
    \label{fig:bivariate_generated}
\end{figure*}

For a higher-level view of each method's ability to capture the unconditional data distribution, we first compare the altitude-averaged wind velocity samples generated by the GMM, DDPM and FM models in \autoref{fig:bivariate_generated}. Here, bivariate plots of the generated $u$ and $v$ velocity components are shown along with the marginal probability density for each component. The true data's marginal densities are also plotted to give perspective on each model's prediction quality. Comparing the generated distributions in \autoref{fig:bivariate_generated} to the true measurement distribution in \autoref{fig:real_bivariate}, it can be seen that the GMM (left) provides a nearly indistinguishable fit to both the joint and marginal distributions, correctly capturing the individual modes in the data. 
By contrast, the DDPM and FM results shown in \autoref{fig:bivariate_generated} (middle) and \autoref{fig:bivariate_generated} (right), respectively, indicating a slightly less appropriate data fit compared to that of the GMM as evident by the mild departure from the real marginal densities for both models.

\begin{figure*}
    \centering
    \includegraphics[width=\linewidth]{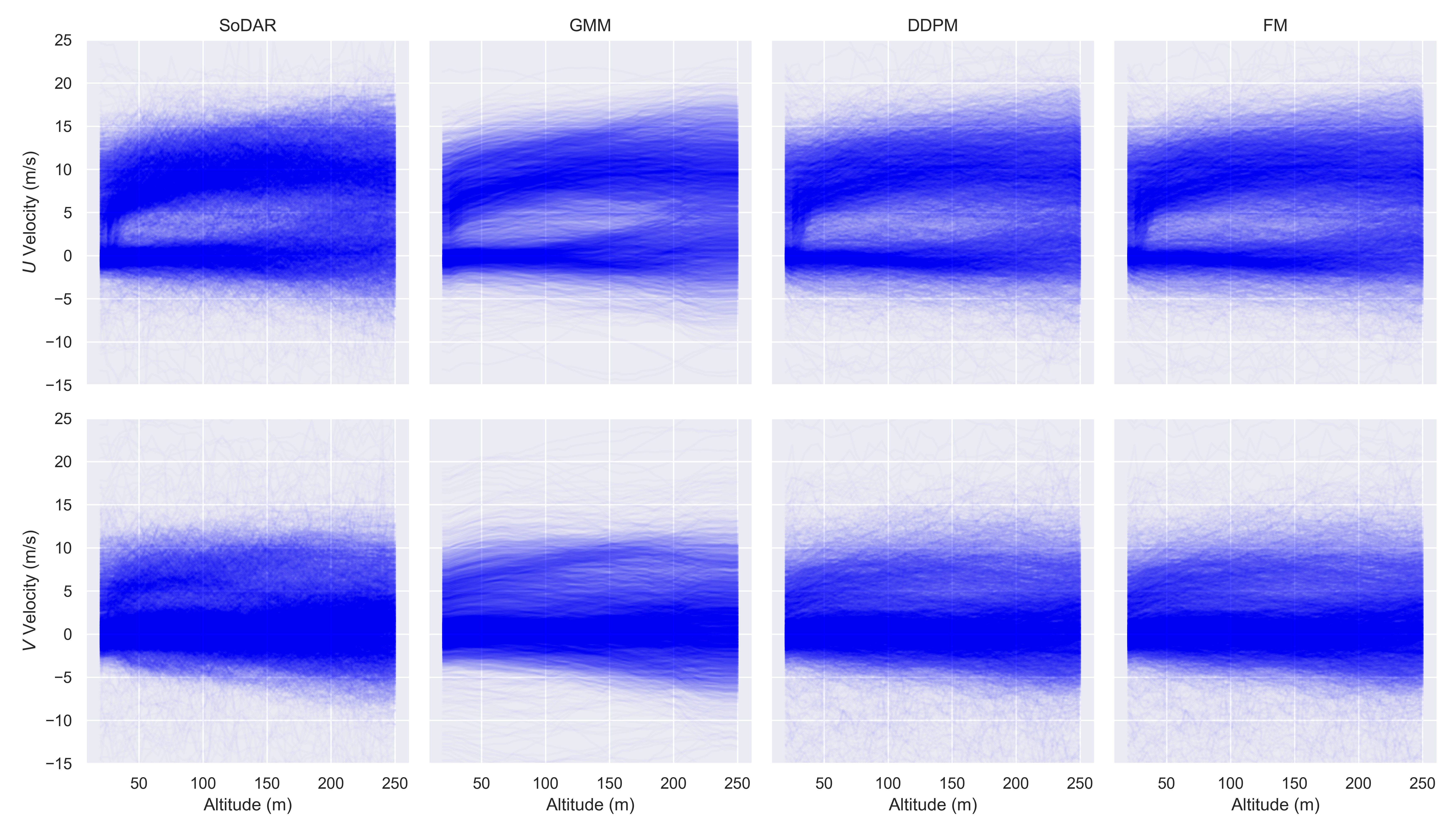}
    \caption{
        \textbf{Samples across the full altitude profile for both $u$ (top) and $v$ (bottom) components from the true SoDAR data, GMM, DDPM and FM models as seen from left to right. 
        The dimension reduction step in the GMM in which we remove some of the high-resolution information can be seen in the smoothness of the data generated from the fitted model. 
        By contrast, both the DDPM and FM models much more faithfully capture the high-resolution information the the profile measurements.}}
    \label{fig:full_wind_realizations}
\end{figure*}

As a more complete visual comparison of each generative model's representation of the true measurement data, realizations of full wind velocities as a function of altitude are presented in \autoref{fig:full_wind_realizations}. 
Although the GMM appeared to provide a slightly better data fit than both DDPM and FM in the altitude-averaged case, the lower model complexity of the GMM results in poor sample realism as shown in \autoref{fig:full_wind_realizations}.
Namely, it can be seen that the full altitude realizations of the GMM are significantly smoother across altitudes than those seen in the true data.
This smoothness is expected given the data reduction step, as higher resolution information is intentionally dropped in favor of computational tractability. 
By contrast, samples drawn from both the DDPM and FM models more faithfully capture the high-resolution information across altitudes and thus provide more visually realistic samples when compared against those from the GMM. 

\begin{figure}
    \centering
    \includegraphics[width=\linewidth]{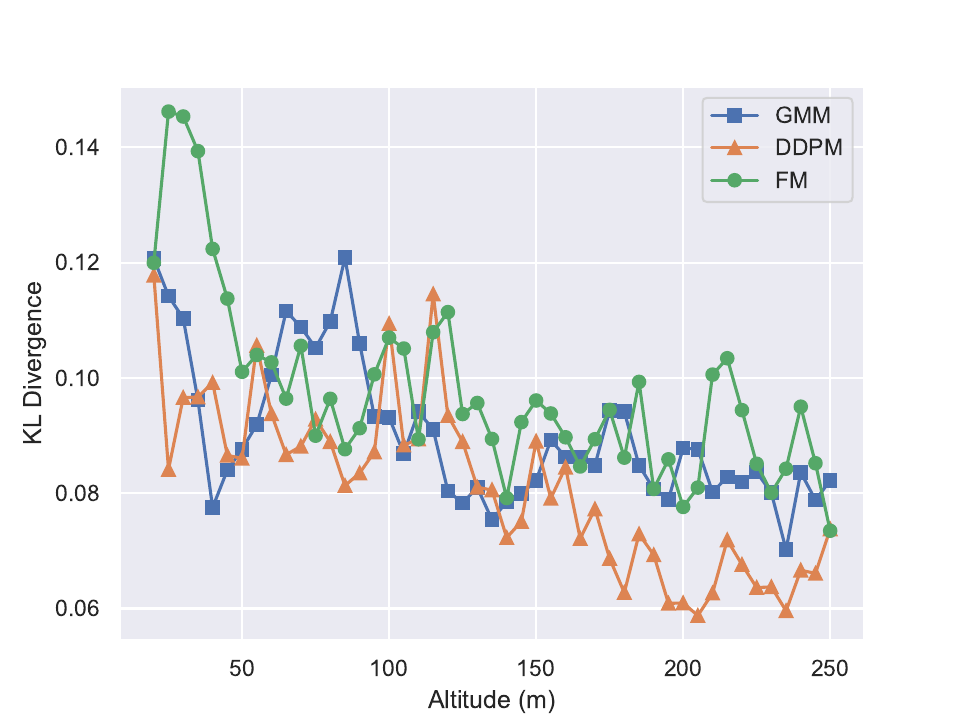}
    \caption{\textbf{Empirical KL-divergence for fitted GMM, DDPM and FM models with the true data across the observed altitudes.}}
    \label{fig:emp_kl_altitude}
\end{figure}

The KL divergence between each method's generated wind velocity distribution versus the measurement distribution as a function of altitude is shown in \autoref{fig:emp_kl_altitude}, providing a quantitative performance comparison of each generative model.
Despite the lack of sample realism from the GMM in \autoref{fig:full_wind_realizations} its ability to match the data distribution across altitudes is quite good in comparison to the DGMs as suggested by \autoref{fig:emp_kl_altitude}. 
Only at the highest altitudes does the DDPM algorithm appear to generate samples which match the measurement distribution better than GMM and FM.

Interestingly, all three models' KL divergences appear to improve as a function of altitude, suggesting that lower altitude microweather distributions are more difficult to represent.

\subsection{Conditional Sampling}

The ability of each model to generate \textit{conditional} wind velocity samples tailored to particular macroweather conditions is explored. 
In our work, we consider two conditioning variables: forecasted wind speed and wind direction. 
Both variables are treated as categorical variables with discrete rather than continuous values due to the relatively small size of the training dataset considered. 
Wind speed is binned into four groups (categories),  $(0.00, 2.23) m/s$, $(2.23, 5.36) m/s$, $(5.36, 8.05) m/s$, $(8.05, 15.65) m/s$, selected to roughly contain the same amount of samples, while four wind directions were chosen that accounted for a majority of the measurements, $\{ SW, W, WNW, WSW \}$.

The DGM models (DDPM and FM) allow for straightforward incorporation of conditional information by taking in macroweather conditions as an extra input in training and sample generation.
For the GMM the process of conditional sample generation is more cumbersome.
A large number of samples must be generated by the GMM, then samples with the desired macroweather conditioning are filtered out through the rejection sampling procedure described in \autoref{sec:gmm_background}. 

\begin{figure*}
    \centering
    \includegraphics[width=\linewidth]{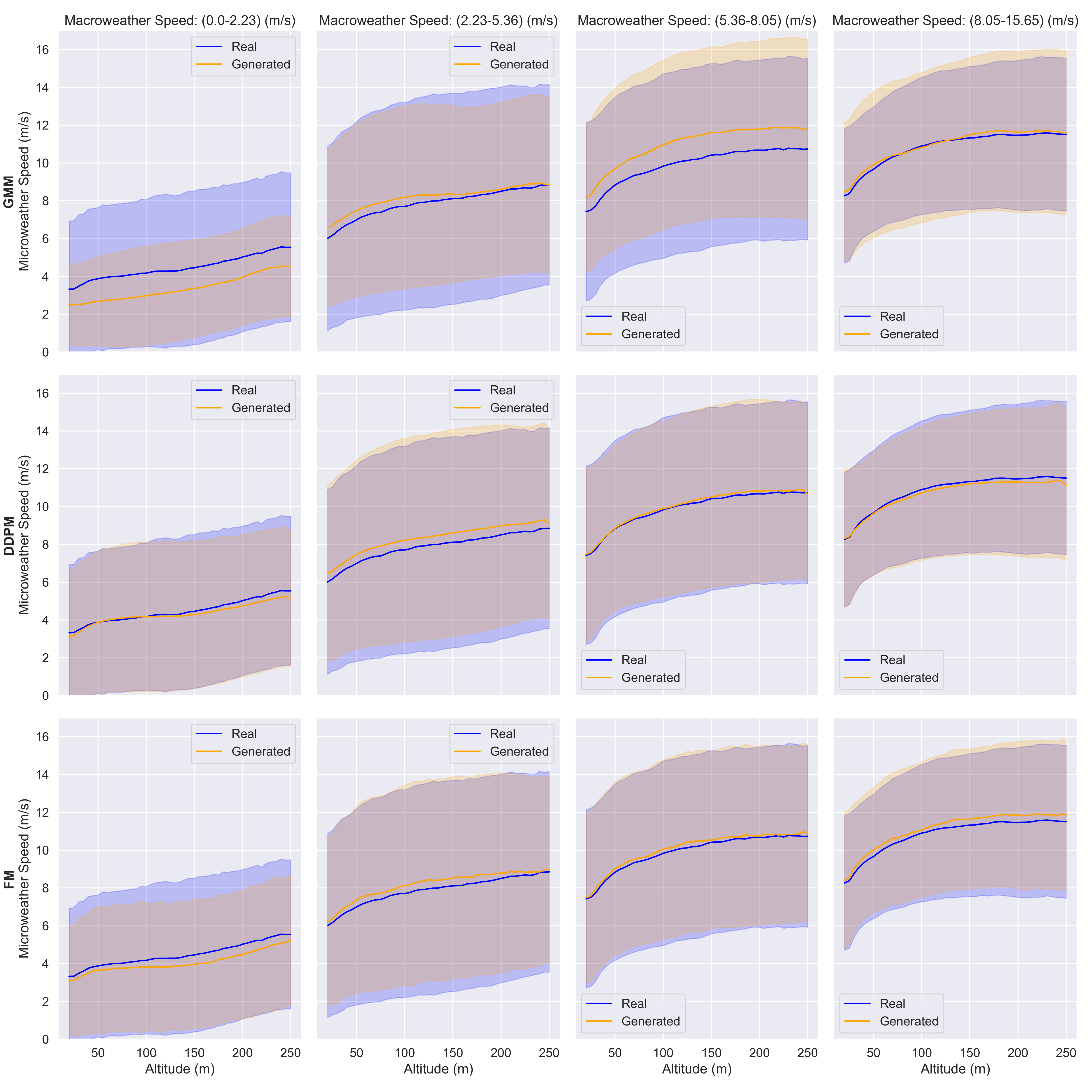}
    \caption{
        \textbf{Mean (solid lines) and variance (shaded region) of the generated distributions by the models compared to the data distribution. 
        The GMM samples on the top row struggle to represent the data at low macroweather wind speeds, however at the highest wind speed the GMM provides a good fit.
        The second row of plots show the DDPM generated samples which share a similar mean and variance to the true data. 
        Similarly, on the third row, the FM generated samples also fit the data well.
        }}
    \label{fig:model_macro_micro_speed}
\end{figure*}

The ability of the generative models to produce wind velocity samples conditioned on macroweather wind speed only is demonstrated in \autoref{fig:model_macro_micro_speed}. 
The mean and standard deviation is computed on generated samples from each model and compared to the real mean and standard deviation across all altitudes in \autoref{fig:model_macro_micro_speed}. 
In this figure, each row denotes a different method and each column denotes a different macroweather wind speed category, with increasing speed from left to right. 
Qualitatively, it can be seen that the GMM struggles to capture the true conditional distributions relative to the DDPM and FM, particularly for the lower speed conditions.
Overall, the DGM appear to follow the true altitude-wise mean and standard deviation of the wind speed well across all four conditional distributions.

\begin{figure*}
    \centering
    \includegraphics[width=\linewidth]{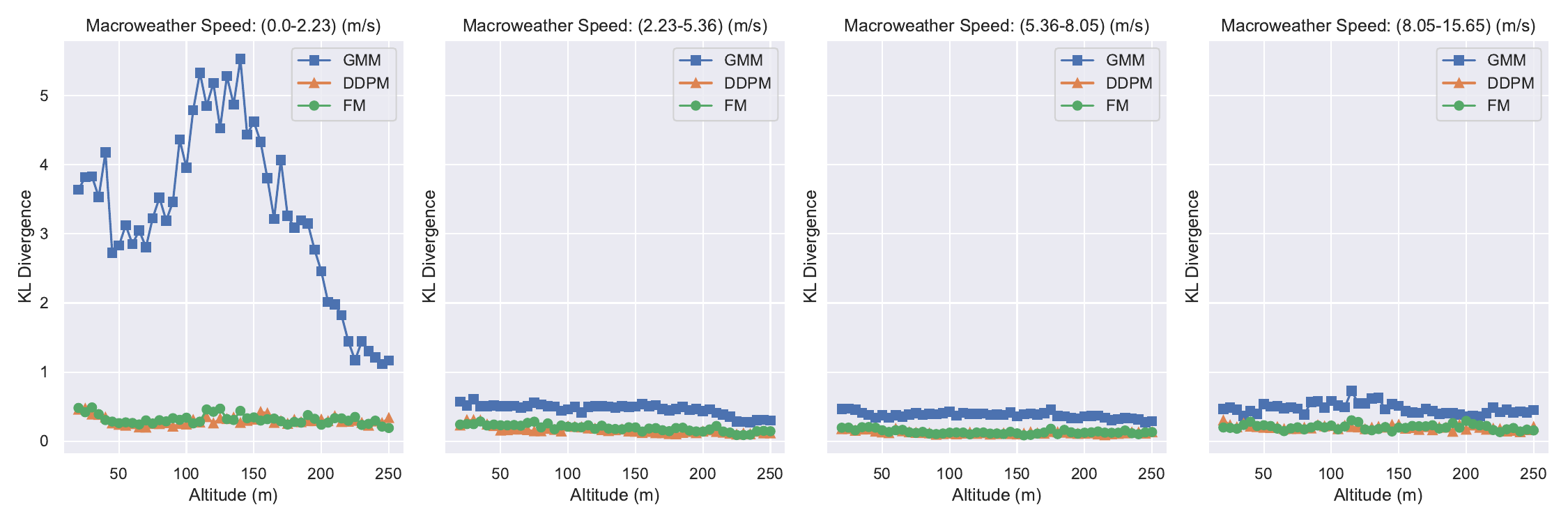}
    \caption{
       \textbf{ KL divergence by altitude for conditional macroweather distributions generated by the models. 
        Due to the GMM's inability to fit to the $\mathbf{(0.0, 2.23)\mbox{ m/s}}$ speed bracket some NaN values were produced because it generated samples with very low likelihood under the data distribution. 
        These nan values were filtered out in order to adequately visualize the results in this plot.}}
    \label{fig:conditional_kl_div_altitude_symmetric}
\end{figure*}

Quantitative evidence of the distribution modeling capabilities of each method is shown in \autoref{fig:conditional_kl_div_altitude_symmetric}.
The KL divergence of the generated distributions are plotted as a function of altitude for each macroweather speed condition.
The GMM has higher KL divergence than the DGMs in each speed conditioning category, particularly in the slowest macroweather wind speed bracket $(0.0, 2.23)\mbox{ m/s}$ a large spike in KL divergence occurs. 
A possible explanation for this behavior is that the microweather distribution for slow macroweather speeds is much different than the other speed brackets.
Additionally, we obsered that the $(0.0, 2.23)\mbox{ m/s}$ bracket contains microweather with more diversity in wind directions whereas higher wind speeds tend to point in a distinct direction as shown in \autoref{fig:bivariate_generated}.
The DGMs are able to maintain low KL divergence across all altitudes in each macro speed range indicating good distribution representation regardless of the microweather characteristics in each bracket.

\begin{figure*}
    \centering
    \includegraphics[width=\linewidth]{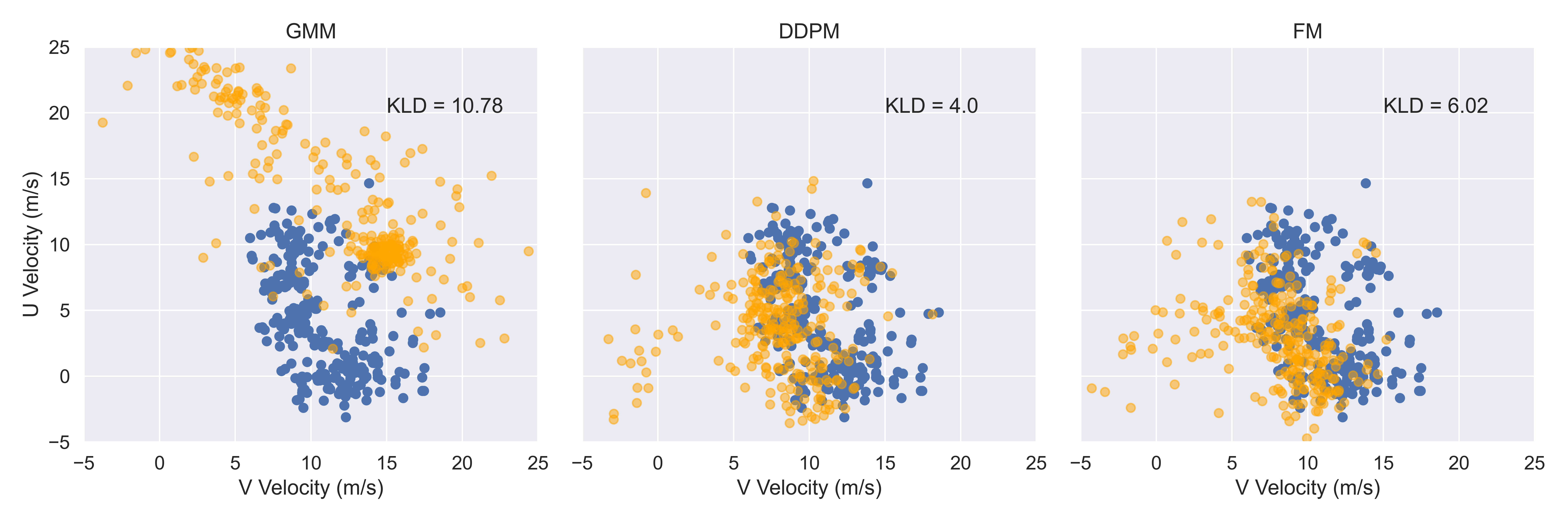}
    \caption{\textbf{Generated samples (orange) compared to real samples (blue) on an unseen combination of macroweather conditions. In this figure the models are trained on all available data except for those corresponding to a macroweather wind speed bracket of $(5.36, 8.05)\mbox{m/s}$ and macroweather wind direction of SW.
    The GMM demonstrates a failure to predict the data distribution while the DGMs both produce samples which better align with the data distribution. The KL divergence (KLD) for each model is indicated.}}
    \label{fig:cherrypicked}
\end{figure*}

Results for producing wind velocity samples conditioned on both macroweather wind speed and direction are now considered. 
Here, the ability of the proposed approach to generalize and predict accurate distributions under combinations of macroweather conditions which are not present in its training dataset is demonstrated. 
This is particularly important for a potential microweather nowcasting system where a temporary wind measurement campaign of a particular location may not be able to fully observe every possible weather interaction, especially when considering continuous conditioning variables. 
To test the ability of the models to generalize effectively to unseen conditions, we perform K-fold cross validation and withhold particular combinations of macroweather conditions during the training of each model. 
At inference time we can evaluate each models' capability to predict the unseen distribution that was withheld from training.

We show a particular example of the generative models' predicted distributions on a particular hold-out macroweather combination (wind speed of (5.36, 8.05)m/s and wind direction of SW) in \autoref{fig:cherrypicked}. 
The GMM in \autoref{fig:cherrypicked} (left) fails to generalize to the hold-out macroweather conditions and predicts a distribution which does not align with the true distribution.
In \autoref{fig:cherrypicked} we selected a particular macroweather combination for which the GMM is capable of generating samples. 
However, in some cases shown in \autoref{fig:gmm_k_fold} in the appendix the GMM fails to allocate probability mass to the hold-out condition combination and therefore does not generate any samples even after applying the rejection sampling procedure $\sim10^8$ times. 
Despite not having access to the specific macroweather combination during training the DGM models are able to correctly identify the shape and location of the data distribution.
There are some outliers in the DDPM samples in \autoref{fig:cherrypicked} (middle) and FM samples in \autoref{fig:cherrypicked} (right), however they are mostly clustered around the correct location. 
Furthermore, unlike the GMM, the DGM are able to guarantee the ability to generate samples in any combination of hold-out macroweather conditions. See the appendix, \autoref{sec:appendix} for detailed results of all 16 macro weather speed-direction conditioning combinations for all three methods.

The results we have demonstrated suggests that the DGM are able to learn how different combinations of macroweather conditions interact with one another in training examples. 
On an unseen pair of conditions the models are able to make a relatively accurate inference on how those conditions will interact. 
This result is consistent with recent literature which has demonstrated that DGMs can learn interactions between different concepts in natural language to synthesize distributions over images which have not been seen during training \cite{esser2024scaling}, \cite{kumari2023multi}, \cite{ramesh2021zero}.
These works also suggest the scalability of DGM approaches to much higher resolution data with a larger number of conditioning variables, including those with continuous rather than categorical values.

\section{Conclusion} \label{sec:conclusions}
This paper presented a generative modeling approach for characterizing local wind velocities based on measurement data. 
It was demonstrated that a probabilistic macro-to-microweather mapping can be learned to produce statistically consistent samples of (microweather) wind velocity vs. altitude that are conditioned on the current (macroweather) forecast for the region. 
A proof of concept was implemented using a dataset comprised of SoDAR wind profiles and weather station data over the same 10 day time period. 
The proposed macro-to-microweather conditional model was implemented using both DDPM and FM, representing state-of-the-art in generative AI, as well as a GMM for a simpler and well-established baseline. 

All three generative models were shown to perform well for unconditional sample generation, producing synthetic wind vs. altitude samples that were statistically consistent with the measurement dataset. 
For generating wind vs. altitude samples tailored to specific weather forecasts with conditional sampling, the DGMs showed superior performance in terms of agreement with the original dataset. 

Furthermore, the GMM was unable to generalize well to hold-out data of macroweather conditions. 
Due to the cumbersome rejection sampling procedure required for the GMM and its lower quality generated samples we conclude that it would not make the ideal candidate for a the development of a microweather nowcasting system. 
Furthermore, it is expected that DGMs would scale more effectively to larger datasets and higher dimensional measurements relative to GMMs.

The proposed work contributes to an important area of research aimed at characterizing wind flows in microenvironments such as those in urban areas. 
The ability to do so is critical for the commercial viability of UAM, which will not gain mainstream adoption until it is perceived to be as safe, reliable, and dependable as other transportation options. 
Relative to existing approaches for assessing microweather winds, the proposed macro-to-microweather generative model could have some practical advantages for UAM:  1) it does not require permanent wind profiler systems (versus field measurement or flow reconstruction approaches), 2) it is computationally-efficient (relative to physics-based modeling), 3) it is probabilistic and able to capture random wind variability (versus deterministic modeling approaches). 
Although it was not explicitly demonstrated here, the approach should also be able to scale well to significantly larger measurement datasets and increased density of spatial/temporal sensor readings by leveraging the scalability of deep neural networks.

A primary limitation of the proposed approach is that predictions of wind velocity are limited to locations where sensor measurements are available. 
This is in contrast to CFD approaches (or data-driven models trained from CFD data) that are able to make continuous predictions in space and time. 
Furthermore, the simple nature of the proof of concept shown here prevented a thorough study on the generality of the method, mainly due to the relatively small measurement dataset used. 
For example, the temporal dependence of the measurement data was ignored in this study due to lack of data. This simplification yields generative models that cannot simulate the evolution of wind fields over time, limiting their practicality for microweather nowcasting or other downstream applications (e.g., UAM performance modeling). 
Similarly, simple categorical conditioning (i.e., bins of forecasted wind speeds) was used in lieu of conditioning on continuous values. 
Finally, the SoDAR measurements considered here were not taken in an urban landscape, likely representing a more straightforward dataset to apply generative modeling to.

These limitations highlight avenues for future work that can be pursued to improve, and more thoroughly demonstrate, the practical advantages of the proposed approach. 
Algorithmic extensions can be explored to enable the generative modeling of full wind velocity fields beyond where sparse measurements are available. 
Here, ideas from operator learning of continuous functions \cite{lu2021learning} as well as approaches to incorporate statistical \cite{wu2020enforcing} and/or physical constraints \cite{RAISSI2019686} into the learning process may potentially allow for inference beyond sensor locations. 
Additionally, an application of the method to a significantly larger dataset from a new NASA measurement campaign is planned. 
It is expected that the new dataset will allow for generative modeling of winds in both space and time, as well as conditioning on weather forecasts as continuous (rather than categorical) variables.

\acknowledgments
The authors would like to thank the Langley Transformation Initiative (LTI) and the Internal Research and Development (IRAD) programs at NASA Langley Research Center for providing funding support for this work.

\bibliographystyle{IEEEtran}
\bibliography{ieee_aerospace_2025}

\newpage
\onecolumn

\begin{appendices}
    
\section{Full K-Fold Cross Validation Results}\label{sec:appendix}

\begin{figure*}[h!]
    \centering
    \includegraphics[width=5in]{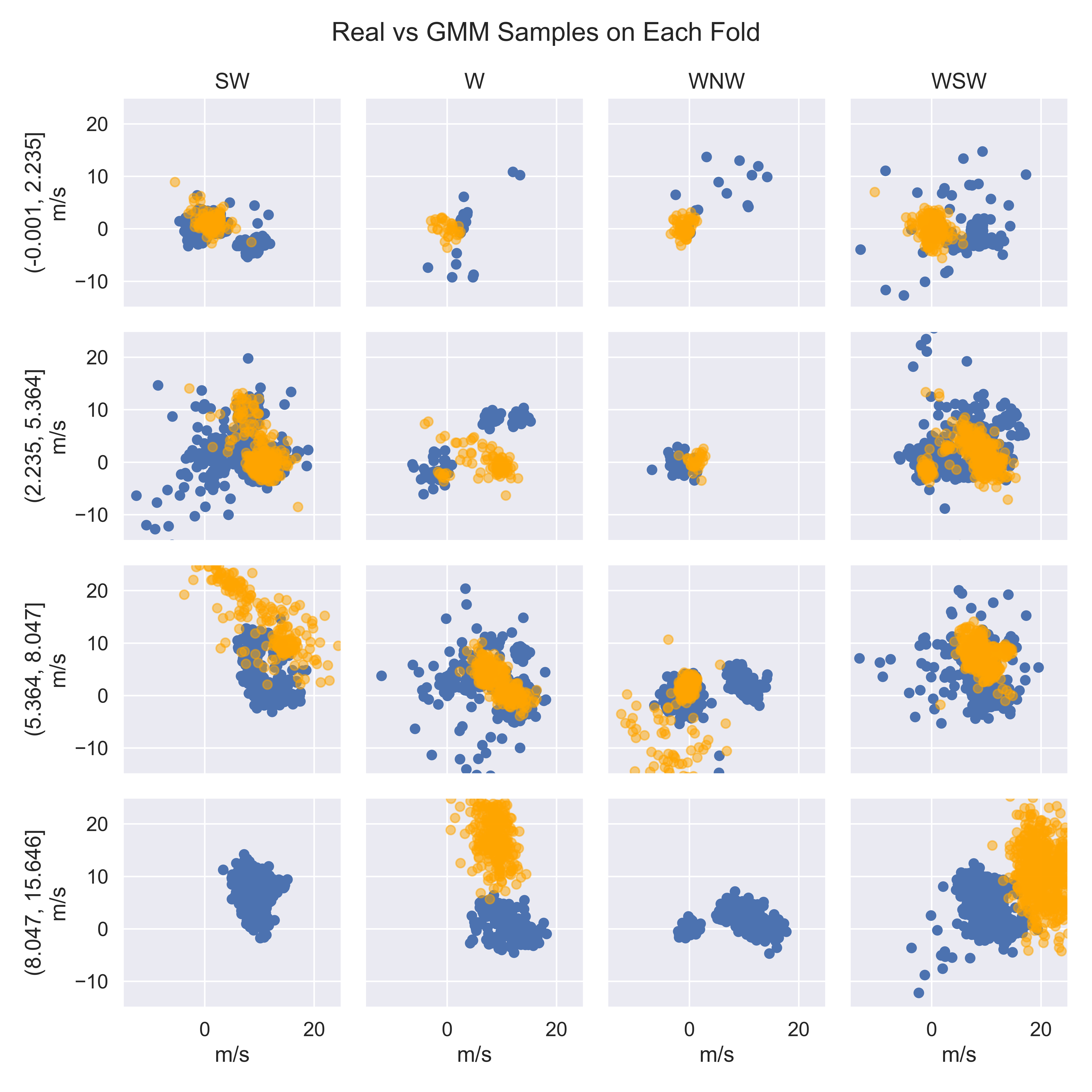}
    \caption{
        \textbf{Each fold of the K-fold cross validation procedure used to test generalzation capabilities of the GMM model.
        Generated samples appear in orange while real samples are blue.
        The predicted distributions appear to be good for certain pairs of macroweather conditions, however, on others the GMM fails.
        Two plots contain no samples from the GMM. In these cases the GMM did not allocate enough probability mass to that particular combination in order to generate samples.
    }}
    \label{fig:gmm_k_fold}
\end{figure*}

\begin{figure*}
    \centering
    \includegraphics[width=5in]{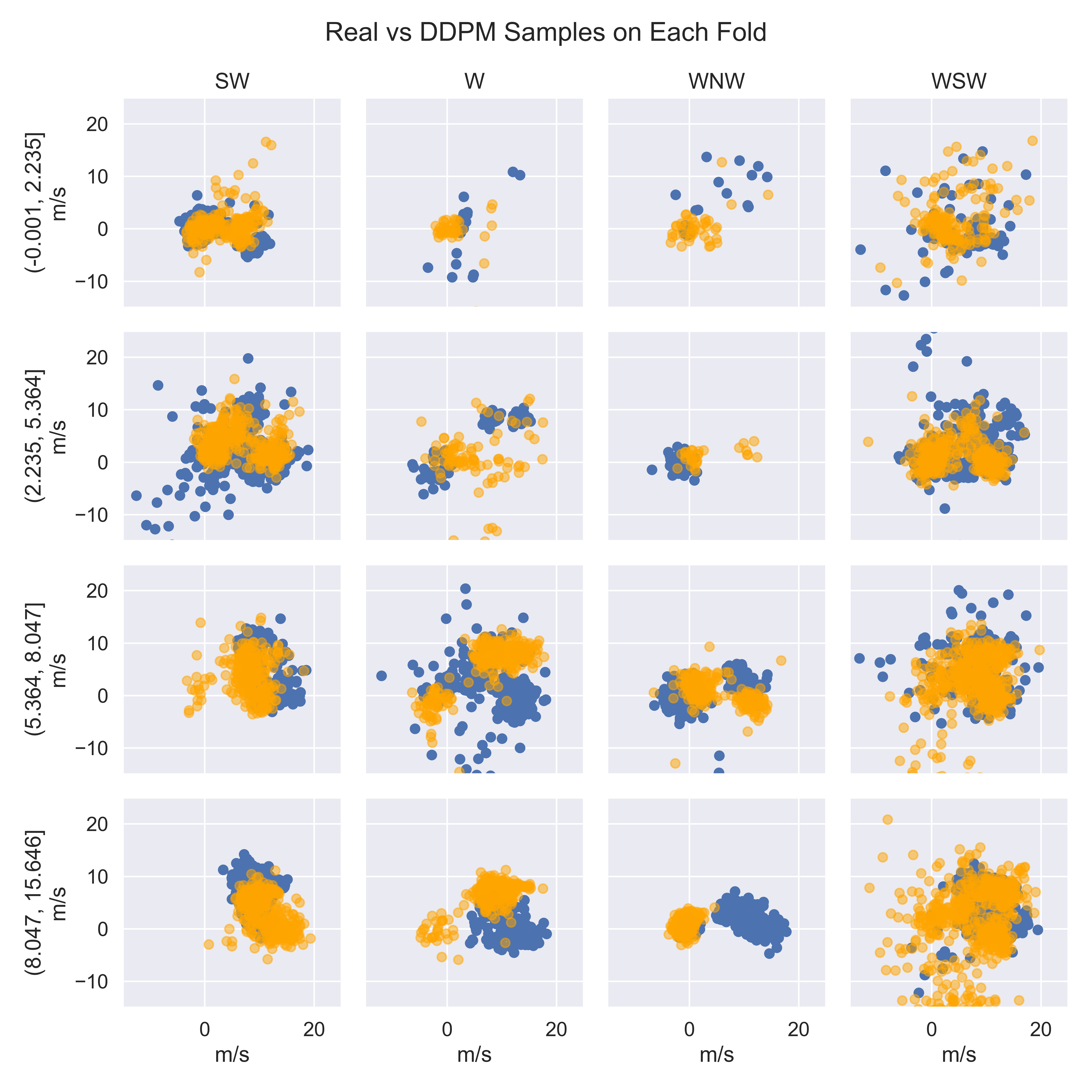}
    \caption{
       \textbf{ Each fold of the K-fold cross validation procedure used to test generalzation capabilities of the DDPM model.
        Generated samples appear in orange while real samples are blue.
        The DDPM is able to generate approximately correct distributions for each pair of macroweather conditions despite not having training data on that pair.
    }}
    \label{fig:ddpm_k_fold}
\end{figure*}

\begin{figure*}
    \centering
    \includegraphics[width=5in]{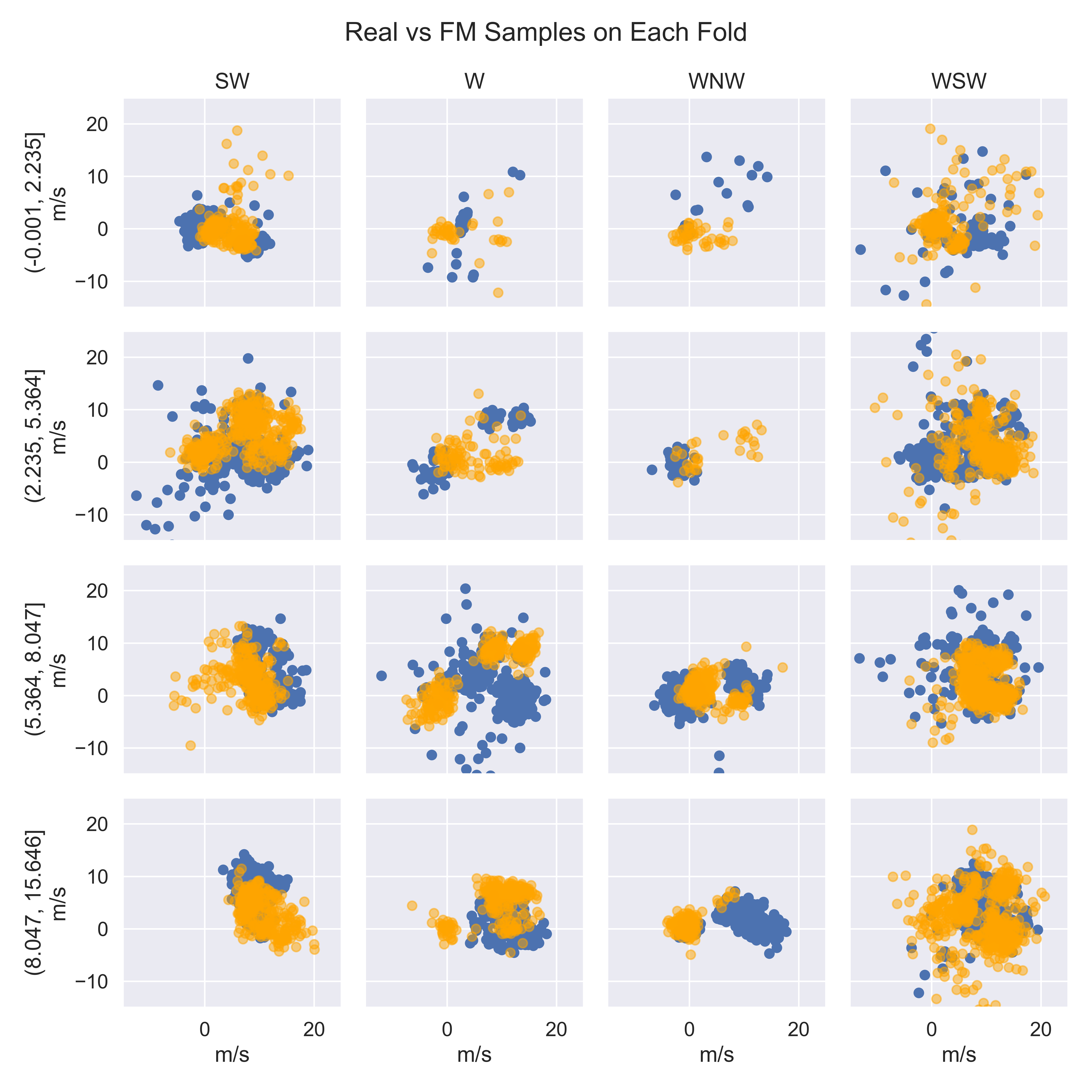}
    \caption{
        \textbf{Each fold of the K-fold cross validation procedure used to test generalzation capabilities of the FM model.
        Generated samples appear in orange while real samples are blue.
        Similar to the DDPM model the FM also generates realistic distributions that match the hold-out data for each fold.}}
    \label{fig:fm_k_fold}
\end{figure*}

This section shows the full set of results for conditional wind velocity generation across all 16 combinations of macroweather wind speed and direction combinations considered. 
Generated samples are compared with measurement data for each conditioning scenario for the GMM in \autoref{fig:gmm_k_fold}, DDPM in \autoref{fig:ddpm_k_fold}, and FM in \autoref{fig:fm_k_fold}. As was illustrated in the representative example in \autoref{fig:cherrypicked}, it can be seen that DDPM and FM generally outperform the GMM in terms of capturing the true conditional distributions. 
Furthermore, there were two conditioning scenarios (directions $SW$ and $WNW$ for speed $(8.05, 15.65)$) where the GMM failed to produce samples using the rejection sampling scheme proposed.

\end{appendices}
\end{document}